\documentstyle[11pt]{article}
\begin{document}
\begin{titlepage}
\begin{flushright}
SUSSEX-AST 94/8-1\\
astro-ph/9408015\\
(August 1994)\\
\end{flushright}
\begin{center}
\Large
{\bf Formalising the Slow-Roll Approximation in Inflation}\\ 
\vspace{.3in}
\normalsize
\large{Andrew R. Liddle, Paul Parsons and John D. Barrow} \\
\normalsize
\vspace{.6 cm}
{\em Astronomy Centre, \\ School of Mathematical and Physical Sciences,\\ 
University of Sussex, \\ Brighton BN1 9QH, U.~K.}\\
\vspace{.6 cm}
\end{center}
\baselineskip=24pt
\begin{abstract}
\noindent
The meaning of the inflationary slow-roll approximation is formalised.
Comparisons are made between an approach based on the Hamilton-Jacobi 
equations, governing the evolution of the Hubble parameter, and 
the usual scenario based on the evolution of the potential energy density. 
The vital role of the inflationary attractor solution is emphasised, and some 
of its properties described. We 
propose a new measure of inflation, based upon contraction of the comoving 
Hubble length as opposed to the usual $e$-foldings of physical expansion, and 
derive relevant formulae. We introduce an infinite hierarchy of slow-roll 
parameters, and show that only a finite number of them are required to 
produce results to a given order. The extension of the slow-roll 
approximation into an analytic slow-roll {\em expansion}, converging on 
the exact solution, is provided. Its role in calculations of inflationary 
dynamics is discussed. We explore rational-approximants as
a method of extending the range of convergence of the slow-roll
expansion up to, and beyond, the end of inflation. 
\end{abstract}

\begin{center}
\vspace{1cm}
PACS number~~~98.80.Cq\\
\vspace*{4cm}
Submitted to {\bf Physical Review D}
\end{center}
\end{titlepage}


\def\theequation{\arabic{section}.\arabic{equation}}

\section{Introduction}

Inflationary universe models are based upon the possibility of slow
evolution of some scalar field $\phi$ in a potential $V(\phi)$
\cite{KT,LL93}. Although some exact solutions of this problem exist,
most detailed studies of inflation have been made using numerical
integration, or by employing an approximation scheme. The `slow-roll
approximation' \cite{ST,SB90,LL92}, which neglects the most slowly changing
terms in the equations of motion, is the most widely used.
Although this approximation works well in many cases, we know that it
must eventually fail if inflation is to end. Moreover, even weak
violations of it can result in significant deviations from the standard
predications for observables such as the spectrum of density
perturbations or the density of gravitational waves in the universe
\cite{SL93,LL93}. As observational data sharpen, it is important to derive a 
suite of predictions for the observables that are as accurate as possible, 
and which cover all possible inflationary models.

In the literature, one finds two different versions of the slow-roll
approximation. The first \cite{ST,LL92} places restrictions on the
form of the potential, and requires the evolution of the
scalar field to have reached its asymptotic form. This approach is most
appropriate when studying inflation in a specific potential. We shall
call it the {\em Potential Slow-Roll Approximation}, or PSRA. The other
form of the approximation places conditions on the evolution of the Hubble 
parameter during
inflation \cite{CKLL}. We call this the {\em Hubble Slow-Roll Approximation}, 
or HSRA. It has distinct advantages over the PSRA, possessing a clearer 
geometrical interpretation and more convenient analytic properties. These 
make it best suited for general studies, where the potential is not 
specified.

In this paper, we clarify the meaning of the different slow-roll 
approximations that exist in the literature, which often describe a variety
of slightly different approximation schemes applied to different
variables at different orders. By formalising the slow-roll
approximation in detail, we will show how to use it as the basis of
a {\em slow-roll expansion} --- a sequence of analytic approximations which
converge to the exact solution of the equations of motion for an
inflationary universe. Such a technique relies strongly on the notion of 
the inflationary attractor, whose properties we describe. The use of Pad\'{e} 
and Canterbury approximants \cite{BAK,PTVF} allows us to further improve the 
range and rate of convergence of this slow-roll expansion.

\section{Equations of Motion and their Solution}
\label{eqM}

We shall deal with the equations of motion in two different forms, both 
appropriate for a homogeneous scalar field $\phi$, evolving in a potential 
$V(\phi)$. We assume that enough inflation has occurred to render the 
densities of all 
other types of matter negligible, and to establish homogeneity in a patch 
at least as big as the horizon. The most familiar form of the equations,
in a zero-curvature Friedmann universe, is
\begin{eqnarray}
\label{eq1}
H^2 & = & \frac{8 \pi}{3 m_{Pl}^2} \left( \frac{1}{2} \dot{\phi}^2 + V(\phi)
	\right) \,, \\
\label{eq2}
\ddot{\phi} + 3 H \dot{\phi} & = & - V' \,,
\end{eqnarray}
where $H \equiv \dot{a}/a$ is the Hubble parameter, $a$ is the scale factor
(synchronous rather than conformal), $m_{Pl}$ the Planck mass, 
overdots indicate derivatives with respect to cosmic 
time $t$, and primes indicate derivatives with respect to the scalar field 
$\phi$.

One can derive a very useful alternative form of these equations by 
using the scalar field as a time variable \cite{M90,SB90,L91}. This
requires that
$\dot{\phi}$ does not change sign during inflation. Without loss of
generality, we can choose $\dot{\phi} > 0$ throughout (this will determine
some signs in later equations). Differentiating Eq.~(\ref{eq1}) with respect 
to $t$ and using Eq.~(\ref{eq2}) gives 
\begin{equation}
\label{aceq}
2 \dot{H} = - \frac{8\pi}{m_{Pl}^2} \dot{\phi}^2 \,.
\end{equation}
We may divide each side by $\dot{\phi}$ to eliminate the time dependence in
the Friedmann equation, obtaining 
\begin{eqnarray}
\label{e1}
\left[ H'(\phi)\right]^2 - \frac{12 \pi}{m_{Pl}^2} H^2(\phi) & = & 
	- \frac{32\pi^2}{m_{Pl}^4} V(\phi) \,,\\
\label{e2} 
\dot{\phi} & = & - \frac{m_{Pl}^2}{4\pi} H'(\phi) \,,
\end{eqnarray}
and Eq.~(\ref{aceq}) implies $\dot{H} \leq 0$. This new set of equations ---
the Hamilton-Jacobi equations --- is normally more convenient than the 
Eqs.~(\ref{eq1}) and (\ref{eq2}). They were used by Salopek and Bond 
\cite{SB90} to establish several important results to which we refer later. 

These equations allow one to generate an 
endless collection of exact inflationary solutions via the following 
procedure \cite{LID90,L91}: 
choose a form of $H(\phi)$, and use Eq.~(\ref{e1}) to find the potential for 
which the exact solution applies; now use Eq.~(\ref{e2}) to find 
$\dot{\phi}$, which allows the $\phi$-dependences to be converted into 
time-dependences, to get $H(t)$; if desired, a further integration gives 
$a(t)$ (though this last step is seldom required). For example, this 
procedure gives a very easy derivation of the exact solution describing 
`intermediate' inflation, which corresponds to the choice $H(\phi) \propto 
\phi^{-\alpha}$ with $\alpha$ a positive constant and $\phi > 0$. This exact 
solution was derived using the Hamilton-Jacobi equations by Muslimov 
\cite{M90} (and 
independently by Barrow \cite{B90} using a different technique). Other fully 
integrated exact solutions have been found by Barrow \cite{BE1,BE2}.

It is normally impossible to make analytic progress by first choosing 
a potential $V(\phi)$, because Eq.~(\ref{e1}) is 
unpleasantly nonlinear. The simplest exception is the exponential potential, 
known to drive power-law inflation, for which the Hamilton-Jacobi 
formalism was used by Salopek and Bond \cite{SB90} to 
find (in parametric form) the general isotropic solution.

\subsection{The potential-slow-roll approximation (PSRA)}

When provided with a potential $V(\phi)$ from which to
construct an inflationary model, the slow-roll approximation is normally 
advertised as requiring the smallness of the two parameters (both 
functions of $\phi$), defined by \cite{LL92}
\begin{eqnarray}
\label{epvd}
\epsilon_{\scriptscriptstyle V}(\phi) & = & \frac{m_{Pl}^2}{16\pi} \left( 
\frac{V'(\phi)}{V(\phi)} 
	\right)^2 \,, \\
\label{etvd}
\eta_{\scriptscriptstyle V}(\phi) & = & \frac{m_{Pl}^2}{8\pi} 
\frac{V''(\phi)}{V(\phi)} \,.
\end{eqnarray}
Henceforth, we refer to them as {\em potential-slow-roll} (PSR) {\em 
parameters}\footnote{To preview what is to come, these parameters are 
sufficient to obtain results to first order in slow-roll. However, the 
general slow-roll expansion requires an infinite hierarchy of parameters 
which will be defined in Section \ref{hier}.}.
Their smallness is used to justify the 
neglect of the kinetic term in the Friedmann equation, Eq.~(\ref{eq1}), and
the acceleration term in the scalar wave equation, Eq.~(\ref{eq2}). 
Unfortunately, the smallness of the PSR parameters is a necessary 
consistency condition, but {\em not} a sufficient one to guarantee that
those terms can be neglected.
The PSR parameters only restrict
the form of the potential, not the 
properties of dynamic solutions. The solutions are more 
general because they possess a freely specifiable parameter, the 
value of $\dot{\phi}$, which governs the size of the kinetic 
term. The 
kinetic term could, therefore, be as large as one wants, regardless of the 
smallness, or otherwise, of these PSR parameters.

In general, this PSR formalism requires a further `assumption'; 
that the scalar field evolves to approach an asymptotic attractor solution, 
determined by
\begin{equation}
\label{ATT}
\dot{\phi} \simeq - \frac{V'}{3H} \,.
\end{equation}
The word `assumption' is placed in quotes here because, in general, 
one is able to test whether Eq.~(\ref{ATT}) is approached for a wide 
range of initial conditions. The inflationary attractor is of vital 
importance in the application of the slow-roll approximation, and we discuss 
its properties in Subsection \ref{Attr}.

\subsection{The Hubble-slow-roll approximation (HSRA)}
\label{hub}

If $H(\phi)$ is taken as the primary quantity, then there is a better
choice of slow-roll parameters. We define the HSR parameters, 
$\epsilon_{{\scriptscriptstyle H}}$
and $\eta_{{\scriptscriptstyle H}}$, by \cite{CKLL}
\begin{eqnarray}
\label{H1}
\epsilon_{\scriptscriptstyle H}(\phi) & = & \frac{m_{Pl}^2}{4\pi} \left( 
\frac{H'(\phi)}{H(\phi)} \right)^2 \,, \\
\label{H2}
\eta_{\scriptscriptstyle H}(\phi) & = & \frac{m_{Pl}^2}{4\pi} 
\frac{H''(\phi)}{H(\phi)} \,.
\end{eqnarray}
These possess an extremely useful set of properties which make them
superior choices to $\epsilon_{{\scriptscriptstyle V}}$ and 
$\eta_{{\scriptscriptstyle V}}$ as descriptors of inflation:
\begin{itemize}
\item We have exactly
\begin{eqnarray}
\epsilon_{\scriptscriptstyle H} & = & 3 \, \frac{\dot{\phi}^2/2}{V + 
\dot{\phi}^2/2} \quad \left( 
= - \frac{d \ln H}{d\ln a} \right) \,, \\
\label{aea}
\eta_{\scriptscriptstyle H} & = & - 3 \, \frac{\ddot{\phi}}{3 H \dot{\phi}} 
\quad 
\left( = - \frac{d 
\ln \dot{\phi}}{d\ln a} = - \frac{d \ln H'}{d \ln a} \right) \,.
\end{eqnarray}
\item $\epsilon_{\scriptscriptstyle H} \ll 1$ is the condition for neglecting 
the first term of 
eq.~(\ref{e1}) [the kinetic term in Eq.~(\ref{eq1})].
\item $|\eta_{\scriptscriptstyle H}| \ll 1$ is the condition for neglecting 
the 
derivative of the first term of
Eq.~(\ref{e1}) [the acceleration term in Eq.~(\ref{eq2})]. As a 
consequence, all the necessary dynamical information is encoded in the HSR 
parameters. They do not need to be supplemented by any assumptions 
about the inflationary attractor, Eq.~(\ref{ATT}).
\item The condition for inflation to occur is {\em precisely}
\begin{equation}
\ddot{a} > 0 \Longleftrightarrow \epsilon_{\scriptscriptstyle H} < 1 \,.
\end{equation}
\end{itemize}
There is an algebraic expression relating $\epsilon_{\scriptscriptstyle V}$ 
to $\epsilon_{\scriptscriptstyle H}$
and $\eta_{\scriptscriptstyle H}$ (using Eq.~(\ref{e1})):
\begin{equation}
\label{epseq}
\epsilon_{\scriptscriptstyle V} = \epsilon_{\scriptscriptstyle H} \left( 
\frac{3-\eta_{\scriptscriptstyle H}}{3-\epsilon_{\scriptscriptstyle H}} 
\right)^2 \,.
\end{equation}
The true endpoint of inflation, gauged by the HSR parameters, occurs 
{\em exactly} at $\epsilon_{{\scriptscriptstyle H}} = 1$.
When using the PSR parameters, this condition is approximate;
inflation ending at $\epsilon_{{\scriptscriptstyle V}} = 1$ is only a 
first-order result.  
 
For $\eta_{\scriptscriptstyle V}$, the relation to the HSR parameters is 
differential rather than algebraic,
\begin{equation}
\label{deta}
\eta_{\scriptscriptstyle V} = \sqrt{\frac{m_{Pl}^2 
\epsilon_{\scriptscriptstyle H}}{4\pi}} \, 
	\frac{\eta_{\scriptscriptstyle H}^{\prime}}{3- 
\epsilon_{\scriptscriptstyle H}} + \left( 
	\frac{3-\eta_{\scriptscriptstyle H}}{3-\epsilon_{\scriptscriptstyle 
H}} \right) \left( \epsilon_H + \eta_H 
	\right) \,;
\end{equation}
although a more compact representation in terms of higher-order
parameters will be presented in Section~\ref{hier}.
This will show that the first term in Eq.~(\ref{deta}) is of higher-order in 
slow-roll, so that 
to lowest-order, one has $\eta_{\scriptscriptstyle V} = 
\eta_{\scriptscriptstyle H} + \epsilon_{\scriptscriptstyle H}$. Note that 
$\eta_{\scriptscriptstyle H}$ and $\eta_{\scriptscriptstyle V}$ are not the 
same to first-order in slow-roll, as one expects from $H^2 
\propto V$. We could have defined $\eta_{\scriptscriptstyle H}$ to coincide 
with $\eta_{\scriptscriptstyle V}$ in 
slow-roll, by defining $\bar{\eta}_{\scriptscriptstyle H} = 
\eta_{\scriptscriptstyle H} - \epsilon_{\scriptscriptstyle H}$,
but we prefer to regard the definitions in Eqs.~(\ref{H1}) and (\ref{H2}) as
fundamental. 

The definitions can be used to derive two useful relations between parameters 
of the same type
\begin{eqnarray}
\label{dife}
\eta_{\scriptscriptstyle H} & = & \epsilon_{\scriptscriptstyle H} - 
\sqrt{\frac{m_{Pl}^2}{16\pi}} \, 
	\frac{\epsilon_{\scriptscriptstyle 
H}^{\prime}}{\sqrt{\epsilon_{\scriptscriptstyle H}}} \,, \\
\label{etp}
\eta_{\scriptscriptstyle V} & = & 2 \epsilon_{\scriptscriptstyle V} - 
\sqrt{\frac{m_{Pl}^2}{16\pi}} \, 
	\frac{\epsilon_{\scriptscriptstyle 
V}^{\prime}}{\sqrt{\epsilon_{\scriptscriptstyle V}}} \,.
\end{eqnarray}
Note that although, as functions, the parameters 
of a given type, either HSR or PSR, are related, their values {\em at a 
given} $\phi$ are independent of one another. One immediately sees the 
different `normalisation' of the $\eta$ from Eqs.~(\ref{dife}) and 
(\ref{etp}).

It is important to stress that although we have derived self-consistent 
exact expressions relating the PSR to the HSR parameters, we cannot invert 
these expressions without first assuming that the evolution has reached the 
attractor, Eq.~(\ref{ATT}). As already mentioned, the attractor constraint 
is part of the structure of the HSRA, but is absent from the PSRA. 
So, while the HSRA implies the PSRA, the converse does not hold 
without assuming the attractor constraint.

\subsection{The inflationary attractor}
\label{Attr}

Already in this Section we have seen how important the notion of the 
inflationary attractor is. The behaviour of this attractor was established by 
Salopek and Bond \cite{SB90}, and we now discuss its properties.

Suppose $H_0(\phi)$ is any solution to the full equation of motion,
Eq.~(\ref{e1}) --- either inflationary or non-inflationary. Consider, first, 
a linear perturbation $\delta H(\phi)$. We shall also assume, and discuss 
further below, that the perturbation does not reverse the sign of 
$\dot{\phi}$. It therefore obeys the linearised equation
\begin{equation}
H_0' \delta H' \simeq \frac{12 \pi}{m_{Pl}^2} H_0 \delta H \,,
\end{equation}
which has the general solution
\begin{equation}
\delta H(\phi) = \delta H(\phi_i) \exp \left( \frac{12 \pi}{m_{Pl}^2} 
\int_{\phi_i}^{\phi} \frac{H_0}{H_0'} d \phi \right) \,.
\end{equation}
Since $H_0'$ and $d\phi$ have, by construction, opposing signs, the integrand 
within the exponential term is negative definite, and hence all linear 
perturbations die away.

If $H_0$ is inflationary, the behaviour is particularly dramatic because the 
condition for inflation bounds the integrand away from zero. Consequently one 
obtains
\begin{equation}
\delta H(\phi) < \delta H(\phi_i) \exp \left( - \frac{6\sqrt{\pi}}{m_{Pl}}
	\left| \phi - \phi_i \right| \right) \,.
\end{equation}
That is, if there is an inflationary solution all linear perturbations 
approach it {\em at least} exponentially fast as the scalar field rolls.

Another way of writing the solution for the perturbation, regardless of 
whether $H_0$ is inflationary or not, is in terms of the amount of expansion 
which occurs, by using the number of $e$-foldings $N$ as defined in the 
following section (Eq.~(\ref{efold})). We get the precise result \cite{SB90}
\begin{equation}
\delta H(\phi) = \delta H(\phi_0) \exp \left( - 3 |N_i - N| \right) \,.
\end{equation}

For non-linear perturbations, the problem is more complex; though all the 
solutions are easily seen to approach each other, we have not shown that they 
do so exponentially quickly. The most awkward case is where a perturbation 
actually reverses the sign of $\dot{\phi}$, as the Hamilton-Jacobi equations 
are singular when that happens. Nevertheless, as long as the perturbation is 
insufficient to knock the scalar field over a maximum in the potential, the 
perturbed solution will inevitably reverse and subsequently pass through 
the initial value $\phi_i$ again; then it can be treated as a perturbation 
with the same sign of $\dot{\phi}$ as the original solution.

The picture that emerges is therefore as follows. Provided the potential is 
able to 
support inflation, the inflationary solutions all rapidly approach one 
another, with exponential rapidity once in the linear regime. Even when 
inflation ends, the universe continues to expand and therefore the solutions 
continue to approach one another. Consequently, even the exit from inflation 
is independent of initial conditions. Note that there is no concept of a 
single `attractor solution': all solutions are attractors for one another and 
converge asymptotically. As we shall see, this is a vital requirement 
if a slow-roll expansion is to make any sense.

The situation where the inflationary attractor does not apply is therefore 
soon after inflation begins. Normally this is in the distant past and of no 
concern. An exception, recently noted \cite{CLLSW}, is hybrid inflation. 
There, the slow-roll parameters rise above unity, halting inflation, and then 
fall back below unity, reaching very small values. Nevertheless, it is easy 
to show that inflation fails to restart despite the smallness of the PSR 
parameters, because there is insufficient time for the solution to approach 
the inflationary attractor. Another similar situation will be discussed in 
Section \ref{conc}.

\section{A Better Measure of Inflation}
\setcounter{equation}{0}

An important quantity for making inflationary predictions is the 
amount of inflation that has taken place. Inflation is commonly 
characterised by the number of $e$-foldings of physical expansion 
that occur, as given by the natural logarithm of the ratio of the final scale 
factor to the initial one. This can be expressed {\em exactly} as,
\begin{equation}
\label{efold}
N \equiv \ln \frac{a_f}{a_i} = - \sqrt{\frac{4\pi}{m_{Pl}^2}}
\int_{\phi_i}^{\phi_f} 
\frac{1}{\sqrt{\epsilon_{{\scriptscriptstyle H}}(\phi)}} \; d\phi \,.
\end{equation}
If one is working in the PSRA then ,provided the 
attractor solution Eq.~(\ref{ATT}) is attained, this may be 
approximated by
\begin{equation}
N(\phi_i, \phi_f) \simeq - \sqrt{\frac{4\pi}{m_{Pl}^2}}
\int_{\phi_i}^{\phi_f}  
\frac{1}{\sqrt{\epsilon_{{\scriptscriptstyle V}}(\phi)}} d\phi \,.
\end{equation}
These formulae are needed to make the connection between horizon-crossing
times in calculations of the production of scalar
and tensor perturbations. A comoving 
scale $k$ crosses outside the Hubble radius at a time which is $N(k)$ 
$e$-foldings from the end of inflation, where
\begin{equation}
\label{Ncross}
N(k) = 62 - \ln \frac{k}{a_0 H_0} - \ln \frac{10^{16} 
        {\rm GeV}}{V_k^{1/4}}
	+ \ln \frac{V_k^{1/4}}{V_{{\rm end}}^{1/4}} - \frac{1}{3} \ln
	\frac{V_{{\rm end}}^{1/4}}{\rho_{{\rm reh}}^{1/4}} \,.
\end{equation}
The subscript `0' indicates present values; 
subscript `k' specifies the value when the wave number 
$k$ crosses the Hubble radius during inflation ($k=aH$);
subscript `end' specifies the value at the end of inflation; and 
$\rho_{{\rm reh}}$ is the energy density of the universe after reheating to 
the standard hot big bang evolution. This calculation assumes that 
instantaneous transitions occur between regimes, and that 
during reheating the universe behaves as if matter-dominated.
Ordinarily, it is taken as a perfectly good approximation that the comoving 
scale presently equal to the Hubble radius crossed outside the Hubble 
radius 60 $e$-foldings from the end of inflation, with all other scales 
relevant to large-scale structure studies following within the next few 
$e$-foldings.

Something like 70 $e$-foldings is normally advertised as the minimum for 
inflation to solve the various cosmological conundrums, such as the flatness 
and horizon problems. However, it is well known that this is an approximation 
(for example in the standard big bang model, the universe expands by more 
than a factor $e^{70}$ between the Planck time and the present, without 
solving the flatness or horizon problems), based on the assumption of a 
constant Hubble parameter. A better measure of inflation is the reduction in 
the size of the comoving Hubble length, $1/aH$. First of all, the condition 
for inflation ($\ddot{a} 
> 0$) is equivalent to ${d[(aH)^{-1}]/dt} < 0$. Secondly, it is  
the reduction of $1/aH$, not that of $1/a$, which solves the flatness and 
horizon problems. And finally, for the generation of perturbations, it is 
the relation of the comoving wavenumber $k$ to $aH$ that is important. We 
therefore define
\begin{equation}
\bar{N} \equiv \ln \frac{(aH)_f}{(aH)_i} \,.
\end{equation}
It can be shown that
\begin{equation}
\label{barN}
\bar{N}(\phi_i, \phi_f) = -\sqrt{\frac{4\pi}{m_{Pl}^2}} 
\int_{\phi_i}^{\phi_f} 
\frac{1}{\sqrt{\epsilon_{{\scriptscriptstyle H}}(\phi)}} \, \left( 1- 
\epsilon_{{\scriptscriptstyle H}}(\phi) 
\right) d\phi \,.
\end{equation}
Since $H$ always decreases, $\bar{N} \leq N$ by definition, with the 
difference indicating the extra amount of expansion, required by the 
decrease of $H$ during inflation. In the extreme slow-roll limit (HSR
or PSR) $N$ and 
$\bar{N}$ coincide. This tells us that the true condition for sufficient 
inflation should be that $\bar{N}$ (not $N$) exceeds 70.

Again, using the potential, we can only write down an approximate relation. 
At lowest-order, $N$ and $\bar{N}$ coincide, while to next order one obtains
\begin{equation}
\label{NBO}
\bar{N}(\phi_i, \phi_f) \simeq - \sqrt{\frac{4\pi}{m_{Pl}^2}}
\int_{\phi_i}^{\phi_f}  
\frac{1}{\sqrt{\epsilon_{{\scriptscriptstyle V}}(\phi)}} \left( 1 - 
\frac{1}{3} 
\epsilon_{{\scriptscriptstyle V}}(\phi) - \frac{1}{3} 
\eta_{{\scriptscriptstyle V}}(\phi) \right) d\phi \,.
\end{equation}

Eq.~(\ref{barN}) gives $\bar{N}$ during an arbitrary inflationary 
epoch (not just quasi-de Sitter).  
It also holds if inflation is interrupted for a period, while 
the dynamics are still dominated by the scalar field. This 
requires $\epsilon_{{\scriptscriptstyle H}} > 1$, causing the integrand in 
Eq.~(\ref{barN}) to 
change sign over a range of $\phi$. This is unlikely, but it has 
recently been noted \cite{RLL} that it arises in a variant of hybrid 
inflation with a quartic potential, where the potential that drives inflation 
is of the form $V \propto (1 + \lambda \phi^4)$. For some values of 
$\lambda$, the potential temporarily steepens sufficiently to suspend 
inflation, while $1/aH$ increases. This must be compensated by 
extra inflationary expansion later on.

To make use of the new expressions, Eqs.~(\ref{barN}) and (\ref{NBO}), 
we need an $\bar{N}(k)$ relation to replace Eq.~(\ref{Ncross}). This is 
simply
\begin{equation}
\label{Nbarcross}
\bar{N}(k) = 62 - \ln \frac{k}{a_0 H_0} - \ln \frac{10^{16} 
{\rm GeV}}{V_{{\rm end}}^{1/4}}  - \frac{1}{3} \ln 
\frac{V_{{\rm end}}^{1/4}}{\rho_{{\rm reh}}^{1/4}} \,.
\end{equation}
Although this strongly resembles Eq.~(\ref{Ncross}), it is in fact slightly
simpler; the difference between $V_k$ and $V_{{\rm end}}$ is now part of the 
definition of $\bar{N}$.

\section{The Hierarchy of Slow-Roll Parameters}
\label{hier}
\setcounter{equation}{0}

We now reconsider the intrinsic structure of the 
slow-roll approximation. The HSR parameters ($\epsilon_{{\scriptscriptstyle 
H}}$ and 
$\eta_{{\scriptscriptstyle H}}$ of Section~\ref{hub}), 
which measure the first and second derivatives of the Hubble parameter, are 
all we require to obtain results to first-order in the slow-roll 
expansion. However, to go beyond this, we require more derivatives, 
necessitating further slow-roll parameters. In general, there will be an 
infinite number of these, incorporating derivatives to all orders; we
shall prove that each additional order in the expansion requires the 
introduction of one new parameter. The formal order of the expansion 
parameter depends on the number of derivatives it contains, two derivatives 
for each order.

In \cite{CKLL2}, where second-order results were derived, the extra
parameter was defined as $\xi_{{\scriptscriptstyle CKLL}} = (m_{Pl}^2/4\pi) 
H'''/H'$, which is of the same order as the others. This definition 
is rather unfortunate; we are meant to be expanding about a flat 
potential, and this parameter is not guaranteed to tend to zero in that 
limit 
because of the derivative in the denominator. This definition has led to some 
confusion (for instance, in \cite{KV} where $\xi$ was not treated as an 
expansion parameter at all). The moral of this is that the hierarchy of 
slow-roll parameters should be carefully defined, so as to tend to zero 
as the potential approaches flatness in arbitrary ways.
Even with this restriction, there are different ways one could define the 
hierarchy. If a superscript $^{(n)}$ indicates the $n$-th derivative with 
respect to $\phi$, the simplest definition would appear to be
\begin{equation}
\label{simp}
\,^{n}\tilde{\beta}_{\scriptscriptstyle H} \equiv \frac{m_{Pl}^2}{4\pi} 
\left( \frac{H^{(n)}}{H} 
\right)^{2/n} \,,
\end{equation}
which gives $^{1}\tilde{\beta}_{\scriptscriptstyle H} \equiv 
\epsilon_{{\scriptscriptstyle H}}$ and 
$^{2}\tilde{\beta}_{\scriptscriptstyle H} 
\equiv \eta_{{\scriptscriptstyle H}}$.
However, there is a superior alternative.

\subsection{The HSR hierarchy}
\label{hpar}

We shall work with the set of definitions
\begin{equation}
\label{HPdef}
\,^{n}\beta_{{\scriptscriptstyle H}} \equiv \left\{\, \prod_{i=1}^{n}\left[- 
\frac{d \ln
H^{\left(i\right)}}{d \ln a}\right] \right\}^{\frac{1}{n}} \,,
\end{equation}
which can be expressed as
\begin{equation}
\label{HPds}
\,^{n}\beta_{{\scriptscriptstyle H}} = \frac{m^{2}_{Pl}}{4\pi} \left(\frac{ 
\left(H'\right)^{n-1}H^{ \left(
n+1\right)}}{ H^{n}}\right)^{\frac{1}{n}} \,.
\end{equation}
These have elegant properties, which we shall use to recast the
HSR approximation as an HSR {\em expansion}. Furthermore, as we 
demonstrate below, only a finite number of these parameters are needed 
to obtain results to any given order. 

It is difficult to incorporate $\epsilon_{{\scriptscriptstyle H}}$ into this
scheme naturally --- it must be defined separately. We use the form
given in Eq.~(\ref{H1}), but shall refer to 
$\epsilon_{{\scriptscriptstyle H}}$ as $\,^{0}\beta_{{\scriptscriptstyle H}}$ 
in later Sections. The first 
parameter this definition yields is $\,^{1}\beta_{{\scriptscriptstyle H}} 
\equiv \eta_{{\scriptscriptstyle H}}$, in accord with Eq.~(\ref{H2}). 
The next four HSR expansion parameters are
\begin{eqnarray}
\label{xiti}
\xi_{{\scriptscriptstyle H}} \equiv \,^{2}\beta_{{\scriptscriptstyle H}} & = 
&  \frac{m^{2}_{Pl}}{4\pi}
\left(\frac{H'H'''}{H^{2}} \right)^{\frac{1}{2}} \,, \\
\sigma_{{\scriptscriptstyle H}} \equiv \,^{3}\beta_{{\scriptscriptstyle H}} & 
= & \frac{m^{2}_{Pl}}{
4\pi}\left(\frac{{H'}^{2}H''''}{H^{3}}\right)^{\frac{1}{3}} \,, \\
\tau_{{\scriptscriptstyle H}} \equiv \,^{4}\beta_{{\scriptscriptstyle H}} & = 
& \frac{m^{2}_{Pl}}
{4\pi}\left(\frac{{H'}^{3}H^{(5)}}{H^{4}}\right)^{\frac{1}{4}} \,, \\
\label{zeq}
\zeta_{{\scriptscriptstyle H}} \equiv \,^{5}\beta_{{\scriptscriptstyle H}} & 
= & \frac{m^{2}_{Pl}}{4\pi}
\left(\frac{{H'}^{4}H^{(6)}}{H^{5}}\right)^{\frac{1}{5}} \,.
\end{eqnarray}

The strength of these definitions is that each parameter combines $H$ and its 
derivatives, raised as a whole to some power $1/n$, where $n \in {\cal
Z}^{\rm +}$. This has the interesting consequence that if we wish to 
convert the Taylor-series of some function of $H(\phi)$, and its derivatives, 
into an expansion in slow-roll parameters, we are guaranteed that the powers 
of any specific parameter, $\,^{n}\beta_{{\scriptscriptstyle H}}$, will be 
integer multiples of $n$. Thus, the lowest-order term we can expect to find 
involving this parameter will be $(^{n}\beta_{{\scriptscriptstyle H}})^{n}$. 
Consequently, if one is interested in expanding this function up to order 
$m$, then at most the first $m+1$ parameters can appear. Thus, despite the 
potentially infinite number of slow-roll parameters, we require only a
finite selection of them at any order.

\subsection{The PSR hierarchy}

In the spirit of the previous subsection, we introduce a hierarchy 
of PSR parameters. As stressed earlier, these are not as useful in 
calculations of inflationary dynamics as the HSR parameters. They only 
classify the flatness of the potential, and so encode no information about 
initial conditions. This necessitates adding the additional attractor 
constraint, Eq.~(\ref{ATT}).

As before, we retain the standard definitions of 
$\epsilon_{{\scriptscriptstyle V}}$ and 
$\eta_{{\scriptscriptstyle V}}$ from Eqs.~(\ref{epvd}) and (\ref{etvd}), and 
encapsulate the 
definition of $\eta_{{\scriptscriptstyle V}}$, and all higher-order 
parameters, in the new set
of quantities
\begin{equation}
\label{Vdef}
\,^{n}\beta_{{\scriptscriptstyle V}} \equiv \frac{m^{2}_{Pl}}{8\pi} \left( 
\frac{d
\ln V}{d \phi}\right) \left\{\,\prod_{i=1}^{n} \left[ \frac{d \ln
V^{(i)}}{d \phi}\right] \right\}^{\frac{1}{n}} \,,
\end{equation}
which reduce to
\begin{equation}
\,^{n}\beta_{{\scriptscriptstyle V}} = \frac{m^{2}_{Pl}}{8\pi} \left( \frac{
\left(V'\right)^{n-1} V^{\left(n+1\right)}}{V^{n}}\right)^{\frac{1}{n}} \,.
\end{equation}
This allows construction of a set of PSR expansion parameters 
\begin{eqnarray}
\xi_{{\scriptscriptstyle V}} \equiv \,^{2}\beta_{{\scriptscriptstyle V}} & = 
&
\frac{m^{2}_{Pl}}{8\pi} \left(\frac{V'V'''}{V^{2}} \right)
^{\frac{1}{2}} \,, \\
\sigma_{{\scriptscriptstyle V}} \equiv \,^{3}\beta_{{\scriptscriptstyle V}} & 
= &
\frac{m^{2}_{Pl}}{8\pi} \left(
\frac{\left(V'\right)^{2}V''''}{V^{3}}\right)^{\frac{1}{3}} \,, \\
\tau_{{\scriptscriptstyle V}} \equiv \,^{4}\beta_{{\scriptscriptstyle V}} & = 
&
\frac{m^{2}_{Pl}}{8\pi} \left(
\frac{\left(V'\right)^{3}V^{\left(5\right)}}{V^{4}}\right)
^{\frac{1}{4}} \,, \\
\zeta_{{\scriptscriptstyle V}} \equiv \,^{5}\beta_{{\scriptscriptstyle V}} & 
= & \frac{m^{2}_{Pl}}{8\pi}
\left(\frac{\left(V'\right)^{4}V^{\left(6\right)}}{V^{5}}
\right) ^{\frac{1}{5}} \,.
\end{eqnarray}

In Eqs.~(\ref{epseq})--(\ref{etp}), we presented some additional
properties of $\epsilon$ and $\eta$. As one may suspect, these can be written 
in terms of the higher-order parameters. An extensive collection are given in 
the appendix; here, we note the exact relations
\begin{eqnarray}
\epsilon_{\scriptscriptstyle V} & = &\epsilon_H \left( 
\frac{3-\eta_{\scriptscriptstyle H}}{3-\epsilon_{\scriptscriptstyle H}} 
\right)^2 \,, \\
\eta_{{\scriptscriptstyle V}} & = & \frac{3\epsilon_{{\scriptscriptstyle H}} 
+ 3\eta_{{\scriptscriptstyle H}} -\eta^{2}_{{\scriptscriptstyle H}} - 
\xi^{2}_{{\scriptscriptstyle H}}}{3 - \epsilon_{{\scriptscriptstyle H}}} \,.
\end{eqnarray}
The latter demonstrates that $\eta_{{\scriptscriptstyle V}} = 
\epsilon_{{\scriptscriptstyle H}} + \eta_{{\scriptscriptstyle H}}$ in the 
lowest-order HSRA, as stated earlier.

\section{From Slow-Roll Approximation to Slow-Roll Expansion}
\setcounter{equation}{0}

A procedure for generating analytic approximations to inflationary solutions, 
from 
a slow-roll approximation, has a broad spectrum of applications. More useful
still is a slow-roll expansion, from which solutions could be generated 
analytically to any required order in the slow-roll approximation. An 
indication of how to go about this was given by Salopek and Bond \cite{SB90}. 
We now show how this may be achieved within the framework of the slow-roll 
parameters.

Let us first emphasise the importance of the attractor behaviour for this
procedure. The general isotropic solution for a given potential possesses one 
free parameter, corresponding to the freedom to specify $H$ (or equivalently 
$\dot{\phi}$) at some initial time. However, the traditional slow-roll 
solution, and its order-by-order corrections that we shall describe, generate 
only a single solution. Unless an attractor exists and has been attained, 
there is no need for this single solution to represent in any way the true
solution for that potential. However, if the attractor has indeed been 
reached, then any particular solution provides an excellent approximation to 
those arising from a wide range of initial conditions. This is particularly 
important when inflation approaches its end, and the slow-roll parameters 
becoming large, because one might na\"{\i}vely assume that the one-parameter 
freedom could be important there. If the attractor solution exists, then 
solutions for a wide range of initial conditions will converge, and 
subsequently all exit inflation in the same way. Hence, an expansion 
approximating one particular solution serves as an excellent approximation to 
them all, provided the initial condition for that solution is not 
pathological (which is prevented by the assumption that energies are less 
than the Planck energy). 

We note that there is a formal problem in attempting to prove `no-hair' 
theorems; if inflation is to end there is no formal asymptotic regime 
\cite{BE1}. However, our requirement is just that enough inflation occurs to 
ensure that the range of values of $\dot{\phi}$ needed to encompass the 
entire family of solutions is sufficiently small to validate the 
Eq.~(\ref{ATT}). In typical models so much inflation occurs that this 
situation is easily achieved.

\subsection{The traditional Taylor-series approach}

We start with a potential $V(\phi)$ for which a solution is desired. 
Typically, we cannot solve exactly for $H(\phi)$; instead, we aim to find 
an approximate solution, in terms of $V(\phi)$ and a multivariate Taylor 
expansion in the PSR parameters, all of which can be computed 
analytically given an analytic $V(\phi)$. First recast Eq.~(\ref{e1}) as
\begin{equation}
\label{Heps}
H^{2}(\phi)  = \frac{8\pi}{{3m_{\rm
Pl}^2}}V(\phi){\left(1-\frac{1}{3}\epsilon_{{\scriptscriptstyle 
H}}(\phi)\right)}^{-1} \,.
\end{equation}
Then seek an approximate solution for $H^{2}(\phi)$ of the form
\begin{equation}
\label{Hser}
H^2(\phi) = \frac{8\pi}{3m^{2}_{Pl}}V(\phi) \left( 1 + a
\epsilon_{{{\scriptscriptstyle V}}}
+ b 
\eta_{{{\scriptscriptstyle V}}} + c \epsilon_{{{\scriptscriptstyle V}}}^2 + d 
\epsilon_{{{\scriptscriptstyle V}}} \eta_{{{\scriptscriptstyle V}}} + 
e \eta_{{{\scriptscriptstyle V}}}^2 + f \xi^{2}_{{\scriptscriptstyle V}} + 
\cdots \right) \,,
\end{equation}
where $a$, $b$,... are constants to be determined. [In fact,
Eq.~(\ref{epseq}) already guarantees that $b=e=f=0$, by forcing every term
in $\epsilon_{{\scriptscriptstyle H}}$ to contain at least one power of 
$\epsilon_{{\scriptscriptstyle V}}$.] Note also that, for reasons
discussed in Section~\ref{hpar}, $\xi^{2}_{{\scriptscriptstyle H}}$ is the 
lowest-order term involving $\xi_{{\scriptscriptstyle H}}$ to appear in 
Eq.~(\ref{Hser}), and 
that no further 
PSR parameters appear at second-order. Thus, we require an expansion of
$\left(1-{\epsilon_{{\scriptscriptstyle H}}}/3\right)^{-1}$ in PSR 
parameters of ascending order. This is most readily achieved by assuming 
general expressions for the HSR parameters, in terms of PSR parameters, with 
a similar form to Eq.~(\ref{Hser}). Starting with general first-order forms, 
these may be substituted into Eq.~(\ref{epseq}), which is then solved for the 
unknown constants by comparing coefficients\footnote{Although the slow-roll 
{\em functions} are all inter-dependent, as all are based on derivatives of 
$V(\phi)$, their values {\em at a given} $\phi$ are independent. 
Consequently, one should imagine that this procedure is being carried out 
separately at each $\phi$. The results are then used to construct 
a function 
of $\phi$.}. After first-order results have been obtained, this
procedure may be repeated iteratively order-by-order. It is a straightforward 
(albeit tedious) matter to invert the expansion 
of $(1-{\epsilon}_{{\scriptscriptstyle H}}/3)$ using the binomial theorem, 
and 
hence obtain the corresponding series for $H^{2}(\phi)$.

We have done this to fourth-order, although the results may be truncated at 
lower-order to obtain more manageable expressions, as desired. We find
\begin{eqnarray}
\label{eps4}
\epsilon_{{\scriptscriptstyle H}} & = & \epsilon_{{\scriptscriptstyle V}} - 
\frac{4}{3}\epsilon^{2}_{{\scriptscriptstyle V}} +
\frac{2}{3}\epsilon_{{\scriptscriptstyle V}}\eta_{{\scriptscriptstyle V}} + 
\frac{32}{9}\epsilon^{3}_{V} +
\frac{5}{9}\epsilon_{V}\eta^{2}_{{\scriptscriptstyle V}} - 
\frac{10}{3}\epsilon^{2}_{{\scriptscriptstyle V}}\eta_{{\scriptscriptstyle 
V}}
+ \frac{2}{9}\epsilon{{\scriptscriptstyle V}}\xi^{2}_{{\scriptscriptstyle V}} 
- \frac{44}{3}\epsilon^{4}_{{\scriptscriptstyle V}} 
+ \frac{530}{27}\epsilon^{3}_{{\scriptscriptstyle 
V}}\eta_{{\scriptscriptstyle V}}\nonumber \\
& & \mbox{}-\frac{62}{9}\epsilon^{2}_{{\scriptscriptstyle 
V}}\eta^{2}_{{\scriptscriptstyle V}} +
\frac{14}{27}\epsilon_{{\scriptscriptstyle V}}\eta^{3}_{{\scriptscriptstyle 
V}} -
\frac{16}{9}\epsilon^{2}_{{\scriptscriptstyle V}}\xi^{2}_{{\scriptscriptstyle 
V}} +
\frac{2}{3}\epsilon_{{\scriptscriptstyle V}}\eta_{{\scriptscriptstyle 
V}}\xi^{2}_{{\scriptscriptstyle V}}
 + \frac{2}{27}\epsilon_{{\scriptscriptstyle 
V}}\sigma^{3}_{{\scriptscriptstyle V}} + {\cal O}_{5} \,.
\end{eqnarray}
The second-order truncation concurs with the result derived in \cite{KV}.

Hence, we find the approximate solution
\begin{eqnarray}
\label{H4}
H^{2}(\phi) & = & \frac{8\pi}{3m_{Pl}^2}V(\phi)\left[1 +
\frac{1}{3}\epsilon_{{\scriptscriptstyle V}} - 
\frac{1}{3}\epsilon^{2}_{{\scriptscriptstyle V}} +
\frac{2}{9}\epsilon_{{\scriptscriptstyle V}}\eta_{{\scriptscriptstyle V}} 
\mbox{} + \frac{25}{27}\epsilon^{3}_{{\scriptscriptstyle V}} + 
\frac{5}{27}\epsilon_{{\scriptscriptstyle V}}\eta^{2}_{{\scriptscriptstyle 
V}} 
 - \frac{26}{27}\epsilon^{2}_{{\scriptscriptstyle 
V}}\eta_{{\scriptscriptstyle V}}\right. \nonumber \\ 
& &\mbox{} + \frac{2}{27}\epsilon_{{\scriptscriptstyle 
V}}\xi^{2}_{{\scriptscriptstyle V}} 
- \frac{327}{81}\epsilon^{4}_{{\scriptscriptstyle V}}
+ \frac{460}{81}\epsilon^{3}_{{\scriptscriptstyle 
V}}\eta_{{\scriptscriptstyle V}} - 
\frac{172}{81}\epsilon^{2}_{{\scriptscriptstyle 
V}}\eta^{2}_{{\scriptscriptstyle V}}
+ \frac{14}{81}\epsilon_{{\scriptscriptstyle V}}\eta^{3}_{{\scriptscriptstyle 
V}}
- \frac{44}{81}\epsilon^{2}_{{\scriptscriptstyle 
V}}\xi^{2}_{{\scriptscriptstyle V}} \nonumber \\
& &\mbox{}+ \frac{2}{9}\epsilon_{{\scriptscriptstyle 
V}}\eta_{{\scriptscriptstyle V}}\xi^{2}_{{\scriptscriptstyle V}}
+ \left.\frac{2}{81}\epsilon_{{\scriptscriptstyle 
V}}\sigma^{3}_{{\scriptscriptstyle V}} + {\cal O}_{5}\right]  \,.
\end{eqnarray}
Thus, we have generated an analytic solution for inflation in the potential
$V(\phi)$, that 
is accurate up to fourth-order in the slow-roll parameters, rather than the 
usual lowest-order.
 
We illustrate this with the specific example of the potential $V(\phi) = m^2 
\phi^2/2$. To second-order in the HSR parameters, one finds
\begin{equation}
\label{epsx}
\epsilon_{{\scriptscriptstyle H}} = \frac{m^{2}_{Pl}}{4\pi \phi^2}\left[1 -
\frac{m^{2}_{Pl}}{6\pi\phi^2} + {\cal
O}\left(\frac{m_{Pl}^4}{\phi^{4}}\right) \right] \,,
\end{equation}
and hence
\begin{equation}
\label{Hx}
H^{2}(\phi) = \frac{4\pi m^{2}\phi^{2}}{3m^{2}_{Pl}}\left[1 +
\frac{m^{2}_{\rm Pl}}{12\pi\phi^{2}} - \frac{m^{4}_{Pl}}{144
\pi^{2}\phi^{4}} + {\cal O}\left(\frac{m_{Pl}^6}{\phi^{6}}\right)
\right] \,.
\end{equation}
Results accurate to one order less, as given in Ref.~\cite{SB90}, can be 
obtained by removing the last
term from both expressions. The 
behaviour of $H(\phi)$ is indicated in Figure 1, 
where for comparison, the exact numerical solution is also shown.

For this particular potential, it is necessary to include both first and 
second-order corrections in order to maintain, as a sensible definition of 
the end of inflation, the condition $\epsilon_{{\scriptscriptstyle H}}(\phi) 
= 1$ (here, we refer to the precise $\epsilon_{{\scriptscriptstyle H}}(\phi)$ 
derived from the approximate solution $H(\phi)$, and not to 
$\epsilon_{{\scriptscriptstyle H}}(\phi)$ truncating to the same order). To 
lowest-order, this is guaranteed because the potential has a minimum where 
$V(\phi) = 0$; but, because corrections become large near the end 
of inflation, they may spoil this. In fact, for a $\phi^2$-potential, 
if one includes only the first-order corrections, they conspire so that 
$\epsilon_{{\scriptscriptstyle H}}(\phi)$ fails to reach unity for any 
$\phi$, despite the solutions being closer to the exact numerical solution 
than the lowest-order results for the bulk of the evolution. Including the 
second-order corrections removes this problem for the $\phi^2$-potential. The 
exact behaviour of $\epsilon_{{\scriptscriptstyle H}} (\phi)$ is shown in 
Figure 2 in each case. For potentials with $\phi^\alpha$ behaviour with 
$\alpha \geq 4$, the end of inflation is well defined, even at first-order.

\subsection{The rational-approximant approach}
\label{paap}

In the chaotic inflationary example presented above, it was seen how, as the 
field rolls toward the minimum of the potential, the Taylor-series 
expansion becomes progressively less accurate due to the dependence of 
the slow-roll parameters on inverse powers of $\phi$. It is not true, in 
general, that $\epsilon_{{\scriptscriptstyle H}}(\phi)$ will be the first 
parameter to become large in this manner. Hence, it is not guaranteed that 
the approximation will describe the evolution through, or even up to, the end 
of inflation. At $\phi = 0$, where $V(\phi) = 0$, the Taylor expansion 
diverges, although inflation necessarily finishes before this.

In cases such as this, rational-approximant techniques \cite{BAK,PTVF}
can be effective. For the single variable case, instead of using a 
single Taylor-polynomial, we approximate using the Pad\'{e} approximant --- a
quotient of {\em two} polynomials --- in the hope of achieving a better range 
and rate of convergence. A particular application of this technique to 
inflation was made in Ref.~\cite{LT94}. Unfortunately, Pad\'{e} approximants 
can only be directly applied to single variable expansions, and we require 
the extension of this theory to multi-variable problems.  

\subsubsection{The Canterbury approximant}

The Canterbury approximant \cite{BAK}
supplements a Pad\'{e} quotient approximant in many variables with a 
minimal Taylor series --- where minimal means containing as few terms as 
possible. So, for a function of $r$ variables $f(x_{1}, x_{2},\cdots, x_{r})$ 
expanded to $n^{th}$-order, we postulate an approximant of the form
\begin{equation}
\label{cant}
\left[ L/M\right]_{f} \equiv \frac{A(x_{1}, x_{2},\ldots , x_{r})}{1+B(x_{1},
x_{2},\ldots , x_{r})} + \sum_{p_{1}=0}^{n}\sum_{p_{2}=0}^{n}\cdots
\sum_{p_{r}=0}^{n}e_{p_{1}p_{2}\ldots
p_{r}}\left[\,\prod_{i=1}^{r} x^{p_{i}}_{i} \right] \,,
\end{equation}
where $A(x_{1}, x_{2}, \ldots, x_{r})$ and $B(x_{1}, x_{2}, \dots,
x_{r})$ are multi-variable polynomials of order $L$ and $M$
respectively, the $e_{p_{1}p_{2}\ldots p_{r}}$ are constants and $p_{i}$ are
the powers of $x_{i}$, ranging between $0$ and $n$. 
The orders of the approximants, $L$ and $M$, are constrained 
by the relation $L + M = n$, and the polynomial $B(x_{1}, x_{2}, \ldots,
x_{r})$ contains no zeroth-order term. The undetermined constants in the 
approximant are fixed by expanding $[L/M]_{f}$ to $n^{th}$-order, and 
matching it with the $n^{th}$-order expansion of $f(x)$. Such an approximant 
often yields faster convergence because the quotient contains better 
estimates of the higher-order terms than a truncated Taylor series. 
Ref.~\cite{BAK} provides a useful introduction to approximant techniques.
 
This approximant is not unique. Any attempt to neglect the Taylor terms in 
Eq.~(\ref{cant}) and solve for the remaining constants fails in general, as 
unfortunately there will be an insufficient number of free constants to do 
this. The choice of which values we assign to the constants in the 
Pad\'{e}-term (and hence the form of the corrective Taylor series), is in 
fact completely arbitrary, demonstrating the non-uniqueness of 
Eq.~(\ref{cant}). We find however, that keeping the Taylor-correction terms 
purely $n^{th}$-order simplifies the analysis somewhat, and also ensures 
that they are small in slow-roll.

We are now in a position to recast the PSRA in the form of Eq.~(\ref{cant}). 
In terms of PSR parameters, during slow-rolling inflation, an arbitrary 
function $f(H(\phi))$ may be approximated by 
\begin{equation}
\left[ L/M\right]_{f} = \frac{A(\epsilon_{{\scriptscriptstyle V}}, 
\eta_{{\scriptscriptstyle V}},\ldots
,\,^{L}\!\beta_{{\scriptscriptstyle V}})}{1+B(\epsilon_{{\scriptscriptstyle 
V}}, \eta_{{\scriptscriptstyle V}},\ldots
,\,^{M}\!\beta_{{\scriptscriptstyle V}})} + 
\sum_{p_{0}=0}^{n}\sum_{p_{1}=0}^{n}\cdots
\sum_{p_{n}=0}^{n}e_{p_{0}p_{1}\ldots p_{n}}
\left[\,\prod_{i=0}^{n} (^{i}\!\beta_{{\scriptscriptstyle V}})^{p_{i}}\right] 
 \,,
\end{equation}
where the $^{n}\beta_{{\scriptscriptstyle V}}$ are the general PSR parameters
defined in Eq.~(\ref{Vdef}); recall, also, that we have
$^{0}\beta_{{\scriptscriptstyle V}} \equiv \epsilon_{{\scriptscriptstyle 
V}}$.

This complicated formalism is clarified by showing it at work to a given 
order. We present a $[2/2]$ Canterbury approximant to $(1 - 
\epsilon_{{\scriptscriptstyle H}}/3)$, and indicate how this may be used to
construct the approximant to $H^{2}(\phi)$. For $(1 - 
\epsilon_{{\scriptscriptstyle H}}/3)$, we obtain
\begin{eqnarray}
\label{ec4}
[2/2]_{\left(1 - \epsilon_{{\scriptscriptstyle H}}/3\right)} & = & \frac{1 +
\frac{21}{4}\epsilon_{{\scriptscriptstyle V}} - 
\frac{7}{3}\eta_{{\scriptscriptstyle V}} -
\frac{53}{18}\epsilon_{{\scriptscriptstyle V}}\eta_{{\scriptscriptstyle V}} + 
\frac{89}{36}\epsilon^{2}_{{\scriptscriptstyle V}} +
\eta^{2}_{{\scriptscriptstyle V}} - \frac{2}{9}\xi^{2}_{{\scriptscriptstyle 
V}}}{1 + \frac{67}{12}\epsilon_{{\scriptscriptstyle V}}
- \frac{7}{3}\eta_{{\scriptscriptstyle V}} - 
\frac{7}{2}\epsilon_{{\scriptscriptstyle V}}\eta_{{\scriptscriptstyle V}} +
\frac{35}{9}\epsilon^{2}_{{\scriptscriptstyle V}} + 
\eta^{2}_{{\scriptscriptstyle V}}
- \frac{2}{9}\xi^{2}_{{\scriptscriptstyle V}}}
-\frac{2}{81}\epsilon_{{\scriptscriptstyle V}}\sigma^{3}_{{\scriptscriptstyle 
V}} \nonumber \\
& & \mbox{}+ \frac{1}{162}\epsilon^{3}_{{\scriptscriptstyle 
V}}\eta_{{\scriptscriptstyle V}} -
\frac{35}{324}\epsilon^{2}_{{\scriptscriptstyle 
V}}\eta^{2}_{{\scriptscriptstyle V}} +
\frac{13}{162}\epsilon^{2}_{{\scriptscriptstyle 
V}}\xi^{2}_{{\scriptscriptstyle V}} +
\frac{1}{27}\epsilon_{{\scriptscriptstyle V}}\eta^{3}_{{\scriptscriptstyle 
V}}  \,.
\end{eqnarray}
 
Diagonal Canterbury approximants (ie, the $[L/L]$ cases), share many of
the useful properties of standard Pad\'{e} approximants ({\em duality},
{\em homographic invariance}, {\em unitarity}, etc), 
and substantially simplify the application of 
the Canterbury technique \cite{BAK}. Use of diagonal approximants, 
where possible, is thus recommended. In particular, the duality property may
be exploited to save considerable effort when calculating
the corresponding expression for $H^{2}(\phi)$ from Eq.~(\ref{ec4}).
The duality property is as follows;
if $f\left( x\right) = \left[g\left( x\right)
\right]^{-1}$ and $g\left(
0\right) {\not=} 0$, then
\begin{equation}
\left[ L/L\right]_{f\left( x\right)} = \left\{\left[
L/L\right]_{g\left(x\right)}\right\}^{-1} \,.
\end{equation}
We may thus obtain a $[2/2]_{H^{2}(\phi)}$ from Eq.~(\ref{ec4}), via
\begin{equation}
\left[ 2/2\right]_{H^{2}(\phi)} = \frac{8\pi}{3m^{2}_{Pl}}
V(\phi)\left\{\left[ 2/2\right]_{\left(1 - \epsilon_{{\scriptscriptstyle 
H}}/3\right)}
\right\}^{-1} \,.
\end{equation}

\subsubsection{Simplified Canterbury approximants}

The Canterbury approximant provides a powerful technique for 
calculations of inflationary dynamics. However, in its full higher-order
glory, it can be quite cumbersome and unwieldy. It is therefore useful to 
find circumstances in which the corrective Taylor series is not required. The 
simplest way to bring this about is to take the $[0/n]$ approximant which, as 
we will show, never needs correcting. However, it may also be true that at 
low orders there is sufficient freedom in the diagonal approximants, due to 
the vanishing of some of the terms in the original Taylor series. 


In fact, the $[1/1]$ approximant has this property. Assuming a general 
form for the $[1/1]_{1-\epsilon_{{\scriptscriptstyle H}}/3}$ approximant, and 
matching to the Taylor expansion for $(1-\epsilon_{{\scriptscriptstyle 
H}}/3)$ from Eq.~(\ref{eps4}) to second-order, yields
\begin{equation}
\label{pex1}
\left[1/1 \right]_{(1 - \epsilon_{{\scriptscriptstyle H}}/3)} \equiv
\frac{a + b\epsilon_{{\scriptscriptstyle V}} + c\eta_{{\scriptscriptstyle 
V}}}{1 +
d\epsilon_{{\scriptscriptstyle V}} + e\eta_{{\scriptscriptstyle V}}} =
1 - \frac{\epsilon_{{\scriptscriptstyle V}}}{3}\left\{1 - 
\frac{4}{3}\epsilon_{{\scriptscriptstyle V}} +
\frac{2}{3}\eta_{{\scriptscriptstyle V}} + {\cal O}_{2} \right\} \,.
\end{equation}
Comparing coefficients, we arrive at the result
\begin{equation}
\label{peps}
\left[1/1\right]_{(1-\epsilon_{{\scriptscriptstyle H}}/3)} = \frac{1 + 
\epsilon_{{\scriptscriptstyle V}} -
\frac{2}{3}\eta_{{\scriptscriptstyle V}}}{1 + 
\frac{4}{3}\epsilon_{{\scriptscriptstyle V}} -
\frac{2}{3}\eta_{{\scriptscriptstyle V}}} \,,
\end{equation}
which, by construction, agrees with the Taylor series to second-order.
The corresponding $H(\phi)$ is easily obtained.
For the $\phi^2$ potential, examined earlier, this does indeed improve on 
the second-order Taylor series given in Eq.~(\ref{Hx}), as shown in 
Figures 1 and 2.

As a final observation, note that if the slow-roll parameters all have the 
same functional form, permitting us to write
\begin{equation}
^{n}\!\beta_{{\scriptscriptstyle V}} = \sum_{i} C_{i} f^{i}(\phi)
\end{equation}
where $C_{i}$ are constants and $f(\phi)$ is an arbitrary function of
$\phi$ (which is small in the PSRA),
then we can always circumvent the need for a corrective Taylor-series by
expanding in powers of $f(\phi)$, thus reducing the problem to the
unmodified Pad\'{e} case.

\section{Conclusions}
\label{conc}
\setcounter{equation}{0}

By defining a suitable hierarchy of parameters, we have extended the 
slow-roll approximation to a slow-roll expansion, allowing 
progressively more accurate {\em analytic} approximations to be
constructed via an order-by-order decomposition in terms of slow-roll 
parameters. The use of rational approximants pushes the range of validity of 
the slow-roll expansion up to, and in many instances beyond, the end 
of inflation. With the accurate observational information becoming available, 
this allows an assessment of the accuracy of calculations within the 
slow-roll approximation, and is especially important with the present 
considerable emphasis focussed on inflationary models which make predictions 
far from the standard (zeroth-order) case.

We have used these parameters to define an improved measure of the amount 
of inflation. However, present uncertainties regarding the physics of 
reheating make it useful only in rather extreme circumstances such as a 
temporary suspension of inflation, during which the universe remains scalar 
field dominated, as in the hybrid inflation model of Ref.~\cite{RLL}.

Let us caution the reader regarding the necessity of the attractor condition 
for the slow-roll expansion to make sense. By incorporating order-by-order 
corrections, we can only generate one solution, $H(\phi)$, out 
of the one-parameter family of actual solutions allowed by the freedom of 
$H$, or equivalently $\dot{\phi}$), permitted by the initial conditions. If 
the attractor hypothesis is not satisfied, then the solution generated --- 
while conceivably an accurate particular solution of the equations of motion 
--- need have no relation to the actual dynamical solutions which might be 
attained. A case in point is the exact `intermediate' inflation solution 
\cite{M90,B90,BL93}. For small $\phi$, this solution corresponds to the 
rather unnatural (and noninflationary) behaviour of the field moving up the 
potential and over a maximum, beyond which 
inflation starts. If one attempts to use our procedure to describe this 
entrance to inflation, the solutions generated bear no particular resemblance 
to the exact solution until well into the inflationary regime\footnote{By 
contrast, the `intermediate solution' can also be employed as the slow-roll 
solution in the simple potential $V \propto \phi^{-\beta}$ (with $\beta$ and 
$\phi$ both positive), where the attractor hypothesis can be applied, though 
the solution to which the expansion process tends would have to be found 
numerically again.}. This serves as a cautionary note, that known exact 
solutions are typically only late-time attractors, and unless a 
significant period of inflation occurs {\em before} the time of interest,  
so that the attractor solution is reached, they are of little relevance.

Importantly, with regard to the exit from inflation, we are on much safer 
ground. It is assumed that enough time has passed for the 
attractor to be reached, and hence all solutions exit from inflation in the 
same way. Therefore, when our expansion procedure supplies a particular 
solution, it provides an excellent description of the way in which the entire 
one parameter family of initial conditions will exit inflation. Without this 
vital point, the generation of solutions via the slow-roll expansion would be 
fruitless.

We have concentrated on the dynamics of inflation, rather than on the 
perturbation spectra produced from them. However, the slow-roll expansion can 
also be brought into play there; as an example, we quote the results for the 
spectral indices $n$ for the density perturbations and $n_T$ for the 
gravitational waves (see \cite{LL93} for precise definitions). These have 
long been taken as approximately $1$ and $0$ respectively; results to 
first-order were given by Liddle and Lyth \cite{LL92} and to second-order by 
Stewart and Lyth \cite{SL93}. With our definitions, these read in the HSRA
and PSRA respectively
\begin{eqnarray}
1-n & = &  4 \epsilon_{{\scriptscriptstyle H}} - 2 \eta_{{\scriptscriptstyle 
H}} + 2(1+c) \epsilon_{{\scriptscriptstyle H}}^2 +
    \frac{1}{2} (3 - 5c) \epsilon_{{\scriptscriptstyle H}} 
\eta_{{\scriptscriptstyle H}} - \frac{1}{2}(3-c)
    \xi_{{\scriptscriptstyle H}}^2 + \cdots \,; \\
    & = &  6 \epsilon_{{\scriptscriptstyle V}} - 2 \eta_{{\scriptscriptstyle 
V}} -\frac{1}{3}(44-18c) \epsilon_{{\scriptscriptstyle V}}^2 
    - (4c-14) \epsilon_{{\scriptscriptstyle V}} \eta_{{\scriptscriptstyle V}} 
- \frac{2}{3} \eta_{{\scriptscriptstyle V}}^2 -\frac{1}{6}
    (13-3c)\xi_{{\scriptscriptstyle V}}^2 \nonumber \\
    & & \mbox{} + \cdots \,; \\
n_T & = &  -2 \epsilon_{{\scriptscriptstyle H}} 
-(3+c)\epsilon_{{\scriptscriptstyle H}}^2 + (1+c) 
\epsilon_{{\scriptscriptstyle H}}
    \eta_{{\scriptscriptstyle H}} + \cdots \,; \\
    & = &  -2 \epsilon_{{\scriptscriptstyle V}} -\frac{1}{3}(8+6c) 
\epsilon_{{\scriptscriptstyle V}}^2 + \frac{1}{3}
    (1+3c) \epsilon_{{\scriptscriptstyle V}} \eta_{{\scriptscriptstyle V}} + 
\cdots \,,
\end{eqnarray}
where $c = 4(\ln 2 + \gamma)$ with $\gamma$ being Euler's constant. Notice 
the factors in $1-n$ change even at first-order, due to the different 
definitions of $\eta$ which have been used. Similarly, we reproduce the 
second-order result for the ratio $R$ of tensor and scalar amplitudes 
\cite{SL93}
\begin{eqnarray}
R & = & \frac{25}{2}\epsilon_{\scriptscriptstyle H} \left[1 + 
2c\left(\epsilon_{\scriptscriptstyle H} - \eta_{\scriptscriptstyle H}
        \right) +\cdots \right] \\
  & = & \frac{25}{2}\epsilon_{\scriptscriptstyle V} \left[1 + 2\left(c - 
\frac{1}{3} \right)
        \left( 2\epsilon_{\scriptscriptstyle V} - \eta_{\scriptscriptstyle V} 
\right) +\cdots \right]
\end{eqnarray}
though it should be noted that this is not a direct observable \cite{LT94}. 
Unlike the relatively simple 
dynamics which we have emphasised in this paper, no way of extending 
these expressions analytically to arbitrary order is known.

\vspace*{12pt}
\noindent
{\em Final note:} As we were completing this paper we received a preprint by 
Lidsey and Waga \cite{LW94} which also discusses the slow-roll approximation, 
although with a considerably different emphasis.

\section*{Acknowledgements}

ARL was supported by SERC and the Royal Society and PP by SERC and
PPARC. ARL would like to thank the Aspen Center for Physics, where this work 
was initiated, for their hospitality, and acknowledges the use of the 
Starlink computer system at the University of Sussex. We thank Andrew 
Laycock, Jim Lidsey, David Lyth and Michael Turner for many helpful 
discussions.
\frenchspacing

\section*{Appendix}
\setcounter{equation}{0}
\def\theequation{A.\arabic{equation}}

We provide here a list of expressions, deemed too cumbersome and obtrusive to 
be imposed upon the main body of text, but which could prove useful in 
certain applications. The first four are exact extensions of 
Eq.~(\ref{epseq}) to higher-order parameters.
\begin{eqnarray}
\label{bigeq}
\eta_{{\scriptscriptstyle V}} & = & \left[ 3 - \epsilon_{{\scriptscriptstyle 
H}}\right]^{-1} ( 
3\epsilon_{{\scriptscriptstyle H}} + 3\eta_{{\scriptscriptstyle H}} 
-\eta^{2}_{{\scriptscriptstyle H}} - \xi^{2}_{{\scriptscriptstyle H}} ) \,, 
\\
\xi^{2}_{{\scriptscriptstyle V}} & = & \left[ 3 - 
\epsilon_{{\scriptscriptstyle H}}\right]^{-2}
(27\epsilon_{{\scriptscriptstyle H}}\eta_{{\scriptscriptstyle H}} + 
9\xi^{2}_{{\scriptscriptstyle H}} - 9\epsilon_{{\scriptscriptstyle 
H}}\eta^{2}_{{\scriptscriptstyle H}} -
12\eta_{{\scriptscriptstyle H}}\xi^{2}_{{\scriptscriptstyle H}} - 
3\sigma^{3}_{{\scriptscriptstyle H}} + 3\eta^{2}_{{\scriptscriptstyle 
H}}\xi^{2}_{{\scriptscriptstyle H}}
+ \eta_{{\scriptscriptstyle H}}\sigma^{3}_{{\scriptscriptstyle H}} ) \,, \\
\sigma^{3}_{{\scriptscriptstyle V}} & = & \left[ 3 - 
\epsilon_{{\scriptscriptstyle H}}\right]^{-3}
( 81\epsilon_{{\scriptscriptstyle H}}\eta^{2}_{{\scriptscriptstyle H}} + 
108\epsilon_{{\scriptscriptstyle H}}\xi^{2}_{{\scriptscriptstyle H}} +
27\sigma^{3}_{{\scriptscriptstyle H}} -54\epsilon_{{\scriptscriptstyle 
H}}\eta^{3}_{{\scriptscriptstyle H}} 
-72\epsilon_{{\scriptscriptstyle H}}\eta_{{\scriptscriptstyle 
H}}\xi^{2}_{{\scriptscriptstyle H}}
-54\eta_{{\scriptscriptstyle H}}\sigma^{3}_{{\scriptscriptstyle H}}\nonumber 
\\
& & \mbox{} -27\xi^{4}_{{\scriptscriptstyle H}} 
-9\tau^{4}_{{\scriptscriptstyle H}}
+9\epsilon_{{\scriptscriptstyle H}}\eta^{4}_{{\scriptscriptstyle H}} 
+12\epsilon_{{\scriptscriptstyle H}}\eta^{2}_{{\scriptscriptstyle 
H}}\xi^{2}_{{\scriptscriptstyle H}}
+18\eta_{{\scriptscriptstyle H}}\xi^{4}_{{\scriptscriptstyle H}} 
+27\eta^{2}_{{\scriptscriptstyle H}}\sigma^{3}_{{\scriptscriptstyle H}} 
+6\eta_{{\scriptscriptstyle H}}\tau^{4}_{{\scriptscriptstyle H}}\nonumber \\
& & \mbox{} -3\eta^{2}_{{\scriptscriptstyle H}}\xi^{4}_{{\scriptscriptstyle 
H}} -4\eta^{3}_{{\scriptscriptstyle H}}\sigma^{3}_{{\scriptscriptstyle H}}
-\eta^{2}_{{\scriptscriptstyle H}}\tau^{4}_{{\scriptscriptstyle H}} ) \,, \\
\tau^{4}_{{\scriptscriptstyle V}} & = & \left[ 3 - 
\epsilon_{{\scriptscriptstyle H}}\right]^{-4} ( 
810\epsilon_{{\scriptscriptstyle H}}\eta_{{\scriptscriptstyle 
H}}\xi^{2}_{{\scriptscriptstyle H}} + 405\epsilon_{{\scriptscriptstyle 
H}}\sigma^{3}_{{\scriptscriptstyle H}}
+81\tau^{4}_{{\scriptscriptstyle H}} 
-810\epsilon_{{\scriptscriptstyle H}}\eta^{2}_{{\scriptscriptstyle 
H}}\xi^{2}_{{\scriptscriptstyle H}}
-405\epsilon_{{\scriptscriptstyle H}}\eta_{{\scriptscriptstyle 
H}}\sigma^{3}_{{\scriptscriptstyle H}}
\nonumber\\
& & \mbox{} -270\xi^{2}_{{\scriptscriptstyle 
H}}\sigma^{3}_{{\scriptscriptstyle H}} -216\eta_{{\scriptscriptstyle 
H}}\tau^{4}_{{\scriptscriptstyle H}} 
-27\zeta^{5}_{{\scriptscriptstyle H}} 
+270\epsilon_{{\scriptscriptstyle H}}\eta^{3}_{{\scriptscriptstyle 
H}}\xi^{2}_{{\scriptscriptstyle H}}
+135\epsilon_{{\scriptscriptstyle H}}\eta^{2}_{{\scriptscriptstyle 
H}}\sigma^{3}_{{\scriptscriptstyle H}} +
270\eta_{{\scriptscriptstyle H}}\xi^{2}_{{\scriptscriptstyle 
H}}\sigma^{3}_{{\scriptscriptstyle H}} \nonumber \\
& & \mbox{}+ 162\eta^{2}_{{\scriptscriptstyle 
H}}\tau^{4}_{{\scriptscriptstyle H}}
+ 27\eta_{{\scriptscriptstyle H}}\zeta^{5}_{{\scriptscriptstyle H}} 
-30\epsilon_{{\scriptscriptstyle H}}\eta^{4}_{{\scriptscriptstyle 
H}}\xi^{2}_{{\scriptscriptstyle H}} 
-15\epsilon_{{\scriptscriptstyle H}}\eta^{3}_{{\scriptscriptstyle 
H}}\sigma^{3}_{{\scriptscriptstyle H}}
-90\eta^{2}_{{\scriptscriptstyle H}}\xi^{2}_{{\scriptscriptstyle 
H}}\sigma^{3}_{{\scriptscriptstyle H}}
-48\eta^{3}_{{\scriptscriptstyle H}}\tau^{4}_{{\scriptscriptstyle H}} 
\nonumber\\
& &\mbox{}-9\eta^{2}_{{\scriptscriptstyle H}}\zeta^{5}_{{\scriptscriptstyle 
H}} 
+10\eta^{3}_{{\scriptscriptstyle H}}\xi^{2}_{{\scriptscriptstyle 
H}}\sigma^{3}_{{\scriptscriptstyle H}} 
+5\eta^{4}_{{\scriptscriptstyle H}}\tau^{4}_{{\scriptscriptstyle H}}
+\eta^{3}_{{\scriptscriptstyle H}}\zeta^{5}_{{\scriptscriptstyle H}} ) \,.
\end{eqnarray}
It is also possible to express any parameter we choose as a first-order
differential relation in terms of lower-order parameters. This was done
in Eqs.~(\ref{dife}) and (\ref{etp}) for $\eta$; we do this
here for $\xi$ and $\sigma$
\begin{eqnarray}
\xi^{2}_{{\scriptscriptstyle H}} & = & \epsilon_{{\scriptscriptstyle 
H}}\eta_{{\scriptscriptstyle H}} -
\sqrt{\frac{m^{2}_{Pl}}{4\pi}}\sqrt{\epsilon_{{\scriptscriptstyle 
H}}}\eta_{{\scriptscriptstyle H}}' \,, \\
\xi^{2}_{{\scriptscriptstyle V}} & = & 2\epsilon_{{\scriptscriptstyle 
V}}\eta_{{\scriptscriptstyle V}} - \sqrt{\frac{m^{2}_{Pl}}
{4\pi}} \sqrt{\epsilon_{{\scriptscriptstyle V}}}\eta_{{\scriptscriptstyle 
V}}' \,, \\
\sigma^{3}_{{\scriptscriptstyle H}} & = & \xi^{2}_{{\scriptscriptstyle H}}
\left(2\epsilon_{{\scriptscriptstyle H}} - \eta_{{\scriptscriptstyle 
H}}\right) -
\sqrt{\frac{m^{2}_{Pl}}{\pi}}
\sqrt{\epsilon_{{\scriptscriptstyle H}}} \xi_{{\scriptscriptstyle 
H}}\xi_{{\scriptscriptstyle H}}' \,, \\
\sigma^{3}_{{\scriptscriptstyle V}} & = & \xi^{2}_{{\scriptscriptstyle V}}
\left(4\epsilon_{{\scriptscriptstyle V}} - \eta_{{\scriptscriptstyle 
V}}\right) -
\sqrt{\frac{m^{2}_{Pl}}{\pi}}
\sqrt{\epsilon_{{\scriptscriptstyle V}}}\xi_{{\scriptscriptstyle V}} 
\xi_{{\scriptscriptstyle V}}' \,.
\end{eqnarray}
These compressions allow us to express any PSR parameter of order $n$ as a
first-order differential relation involving HSR parameters of order not
exceeding $n$, as was done in Eq.(\ref{deta}) for $\eta_{{\scriptscriptstyle 
V}}$. 
We present the case for $\xi_{{\scriptscriptstyle V}}$, although such a
result may be derived for any of the higher-order parameters,
\begin{eqnarray}
\label{liteq}
\xi^{2}_{{\scriptscriptstyle V}} & = & \left[ 3 
-\epsilon_{{\scriptscriptstyle H}}\right]^{-2}\left\{
27\epsilon_{{\scriptscriptstyle H}}\eta_{{\scriptscriptstyle H}} + 
9\xi^{2}_{{\scriptscriptstyle H}} -12\eta_{{\scriptscriptstyle 
H}}\xi^{2}_{{\scriptscriptstyle H}}
-9\epsilon_{{\scriptscriptstyle H}}\eta^{2}_{{\scriptscriptstyle H}} 
+3\eta^{2}_{{\scriptscriptstyle H}}\xi^{2}_{{\scriptscriptstyle H}}\right.
\nonumber\\
& & \left. \mbox{}+\left(\eta_{{\scriptscriptstyle H}} 
-3\right)\xi_{{\scriptscriptstyle H}}\left[\xi_{{\scriptscriptstyle H}}
\left(2\epsilon_{{\scriptscriptstyle H}} -\eta_{{\scriptscriptstyle 
H}}\right)
-\sqrt{\frac{m^{2}_{Pl}}{\pi}}\sqrt{\epsilon_{{\scriptscriptstyle 
H}}}\xi_{{\scriptscriptstyle H}}' \right]\right\} 
\,.
\end{eqnarray}
Eqs.~(\ref{bigeq})--(\ref{liteq}) are all exact. We now 
give some
approximate formulae, inverting some of the above relations to yield
expressions for HSR parameters in terms of PSR parameters. If necessary, 
these can be fitted to Pad\'{e} or Canterbury approximants, using the
methods outlined in Section \ref{paap}. We have already stated the
result for $\epsilon_{{\scriptscriptstyle H}}$ (Eq.~(\ref{eps4})); here we 
give the
higher-order parameters,
\begin{eqnarray}
\label{eta4}
\eta_{{\scriptscriptstyle H}} & = & \eta_{{\scriptscriptstyle V}} - 
\epsilon_{{\scriptscriptstyle V}} 
+\frac{8}{3}\epsilon^{2}_{{\scriptscriptstyle V}}
+\frac{1}{3}\eta^{2}_{{\scriptscriptstyle V}} 
-\frac{8}{3}\epsilon_{{\scriptscriptstyle V}}\eta_{{\scriptscriptstyle V}}
+\frac{1}{3}\xi^{2}_{{\scriptscriptstyle V}} 
-12\epsilon^{3}_{{\scriptscriptstyle V}} 
+\frac{2}{9}\eta^{3}_{{\scriptscriptstyle V}}
+16\epsilon^{2}_{{\scriptscriptstyle V}}\eta_{{\scriptscriptstyle V}} 
\nonumber \\
& & \mbox{} -\frac{46}{9}\epsilon_{{\scriptscriptstyle 
V}}\eta^{2}_{{\scriptscriptstyle V}} -
\frac{17}{9}\epsilon_{{\scriptscriptstyle V}}\xi^{2}_{{\scriptscriptstyle V}} 
+\frac{2}{3}\eta_{{\scriptscriptstyle V}}\xi^{2}_{{\scriptscriptstyle V}}
+\frac{1}{9}\sigma^{3}_{{\scriptscriptstyle V}} +{\cal O}_{4} \,, \\
\xi^{2}_{{\scriptscriptstyle H}} & = & \xi^{2}_{{\scriptscriptstyle V}} 
-3\epsilon_{{\scriptscriptstyle V}}\eta_{{\scriptscriptstyle V}}
+3\epsilon^{2}_{{\scriptscriptstyle V}} -20\epsilon^{3}_{{\scriptscriptstyle 
V}} +26\epsilon^{2}_{{\scriptscriptstyle V}}\eta_{{\scriptscriptstyle V}}
-7\epsilon_{{\scriptscriptstyle V}}\eta^{2}_{{\scriptscriptstyle V}} 
-\frac{13}{3}\epsilon_{{\scriptscriptstyle V}}\xi^{2}_{{\scriptscriptstyle 
V}}
\nonumber \\
& & \mbox{} +\frac{4}{3}\eta_{{\scriptscriptstyle 
V}}\xi^{2}_{{\scriptscriptstyle V}}
+\frac{1}{3}\sigma^{3}_{{\scriptscriptstyle V}} +{\cal O}_{4} \,, \\
\sigma^{3}_{{\scriptscriptstyle H}} & = & \sigma^{3}_{{\scriptscriptstyle V}} 
- 3\epsilon_{{\scriptscriptstyle V}}\eta^{2}_{{\scriptscriptstyle V}}
+18\epsilon^{2}_{{\scriptscriptstyle V}}\eta_{{\scriptscriptstyle V}} 
-15\epsilon^{3}_{{\scriptscriptstyle V}}
-4\epsilon_{{\scriptscriptstyle V}}\xi^{2}_{{\scriptscriptstyle V}} +{\cal 
O}_{4} \,,
\end{eqnarray}
Note that these inversions are only valid when 
the attractor condition, Eq.~(\ref{ATT}), holds. The second-order
truncation of Eq.~(\ref{eta4}) is compatible with the result presented in
\cite{KV}. 
\newpage
\section*{Figure Captions}

\vspace{24pt}
\noindent
{\em Figure 1}\\
A comparison of different analytic approximations with the exact, numerically 
generated, solution $H(\phi)$ for a potential $V(\phi) \propto \phi^2$, near 
the end of inflation. The normalisation of $H$ is arbitrary. The slow-roll 
approximation and its first and second-order corrected versions are shown, 
together with the $[1/1]$ rational approximant introduced in Subsection 
\ref{paap}, which is also a second-order correction. The rational approximant 
performs the best, as is more clearly seen in Figure 2.

\vspace{24pt}
\noindent
{\em Figure 2}\\
The same comparison as Figure 1, but this time showing the exact 
$\epsilon_H(\phi)$ corresponding to each of the solutions. Recall that the 
end of inflation is at $\epsilon_H = 1$. The pathological behaviour of the 
first-order corrected solution, for which inflation never ends in this 
potential, is clear; all other solutions have a satisfactory end to 
inflation, with the rational approximant providing the best overall 
approximation to the exact solution.


\begin{thebibliography}{99}
\bibitem{KT} A. H. Guth, Phys. Rev. D{\bf 23}, 347 (1981). E. W. Kolb and 
	M. S. Turner, {\em The Early Universe}, (Addison-Wesley, Redwood
	City, CA, 1990).
\bibitem{LL93} A. R. Liddle and D. H. Lyth, Phys. Rep. {\bf 231}, 1 (1993).
\bibitem{ST} P. J. Steinhardt and M. S. Turner, Phys. Rev.
	D{\bf 29}, 2162 (1984).
\bibitem{SB90} D. S. Salopek and J. R. Bond, Phys. Rev. D{\bf 42}, 3936
	(1990).
\bibitem{LL92} A. R. Liddle and D. H. Lyth, Phys. Lett. {\bf B291}, 391
	(1992).
\bibitem{SL93} E. D. Stewart and D. H. Lyth, Phys. Lett {\bf B302}, 171
	(1993).
\bibitem{CKLL} E. J. Copeland, E. W. Kolb, A. R. Liddle and J. E. Lidsey,
	Phys Rev D{\bf 48}, 2529 (1993).
\bibitem{BAK} G. A. Baker, jr. and P. Graves - Morris, 
        {\em Encyclopedia of Mathematics and its Applications}, volumes         
        {\bf 13} \& {\bf 14}, Addison-Wesley (1981).
\bibitem{PTVF} W. H. Press, S. A. Teukolsky, W. T.  Vetterling and B. P.
	Flannery, {\it Numerical Recipes} ({\it 2nd edition}) (Cambridge 
	University Press, Cambridge, 1993).
\bibitem{M90} A. G. Muslimov, Class. Quant. Grav {\bf 7}, 231 (1990).
\bibitem{L91} J. E. Lidsey, Phys. Lett. {\bf B273}, 42 (1991).
\bibitem{LID90} J. E. Lidsey, Class. Quant. Grav. {\bf 8}, 923 (1990).
\bibitem{B90} J. D. Barrow, Phys. Lett. {\bf B235}, 40 (1990).
\bibitem{BE1} J. D. Barrow, Phys. Rev. D{\bf 48}, 1585 (1993).
\bibitem{BE2} J. D. Barrow, Phys. Rev. D{\bf 49}, 3055 (1994).
\bibitem{CLLSW} E. J. Copeland, A. R. Liddle, D. H. Lyth, E. D. Stewart and
	D. Wands, Phys. Rev. D{\bf 49}, 6410 (1994).
\bibitem{RLL} D. Roberts, A. R. Liddle and D. H. Lyth, ``False Vacuum
	Inflation with a Quartic Potential'', Sussex preprint (1994).
\bibitem{CKLL2} E. J. Copeland, E. W. Kolb, A. R. Liddle and J. E. Lidsey,
	Phys Rev D{\bf 49}, 1840 (1993).
\bibitem{KV} E. W. Kolb and S. L. Vadas, ``Relating spectral indices to
        tensor and scalar amplitudes in inflation'', Fermilab preprint
       	FERMILAB-Pub/046-A, astro-ph/9403001 (1994). 
\bibitem{LT94} A. R. Liddle and M. S. Turner, Phys. Rev. D{\bf 50}, July 15th
	1994.
\bibitem{BL93} J. D. Barrow and A. R. Liddle, Phys. Rev. D{\bf 47}, R5129
	(1993).
\bibitem{LW94} J. E. Lidsey and I. Waga, ``The Andante Regime of Scalar Field
	Dynamics'', Fermilab preprint Fermilab-Pub-94-223-A (1994).
\end{thebibliography}
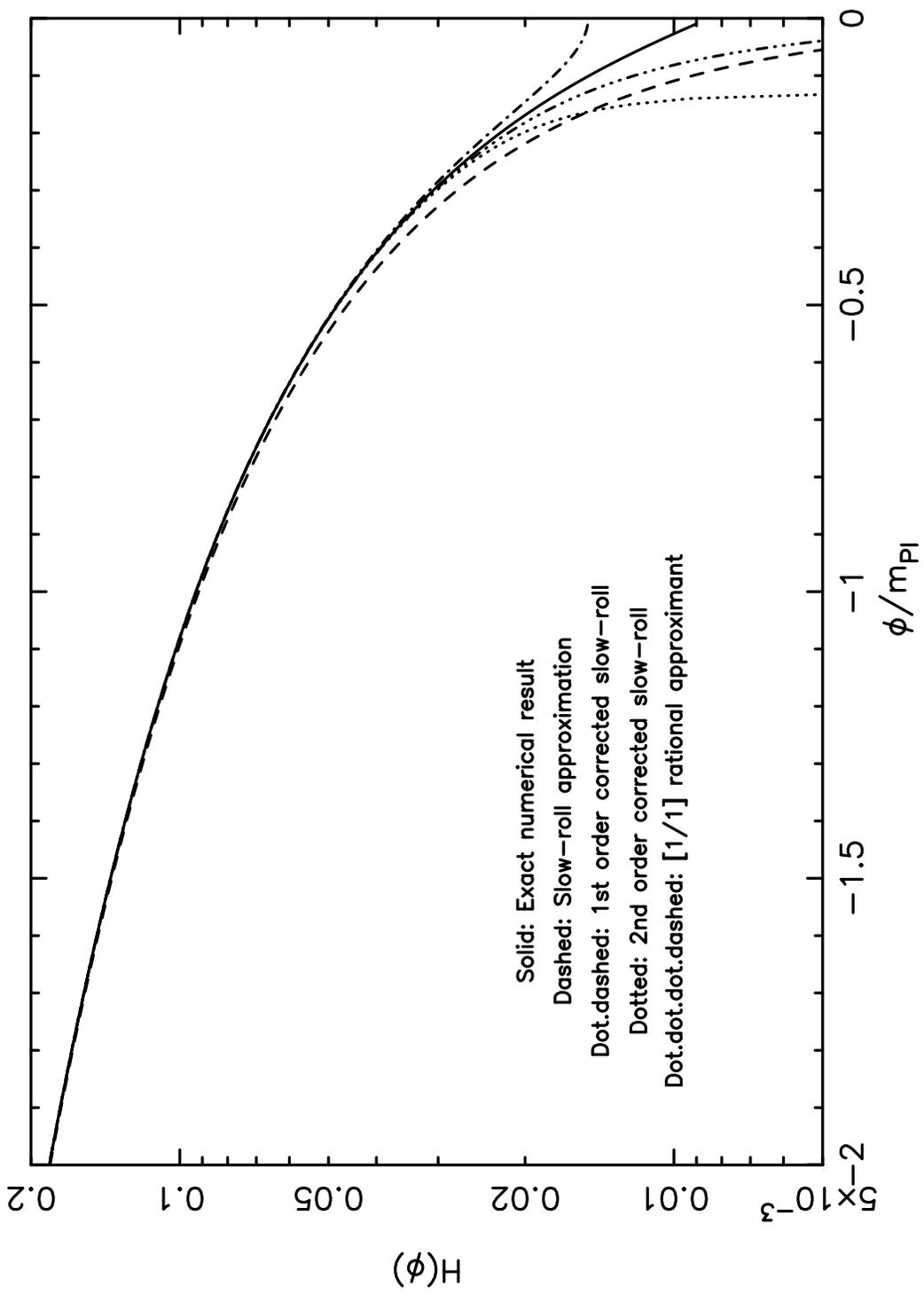
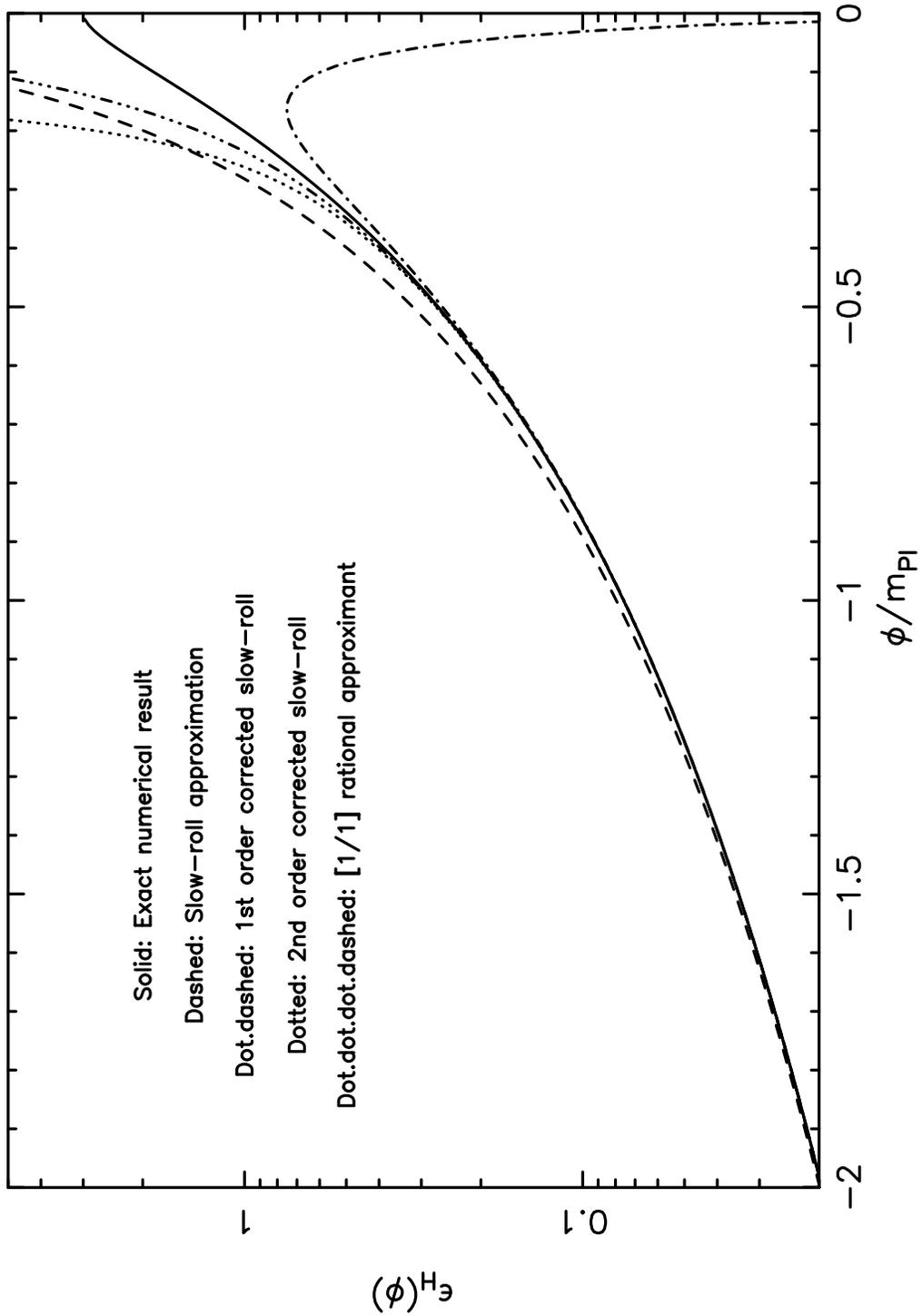
\end{document}